\documentclass[prl,twocolumn,showpacs,amsmath,amssymb]{revtex4}
\usepackage{epsfig}
\usepackage{mathrsfs}

\newcommand{\calA}{\mathcal A}
\newcommand{\calC}{\mathcal C}
\newcommand{\calP}{\mathcal P}
\newcommand{\rrr}{\mathbf{r}}
\newcommand{\nnn}{\mathbf{n}}
\newcommand{\nablabf}{\boldsymbol{\nabla}}
\newcommand{\co}{c^{{}}_0}
\newcommand{\No}{N^{{}}_0}

\begin{document}

\title{Universality in edge-source diffusion dynamics}

\author{Niels Asger Mortensen, Fridolin Okkels, and Henrik Bruus}

\affiliation{MIC -- Department of Micro and Nanotechnology,
NanoDTU, Technical University of Denmark, Bldg.~345 east, DK-2800
Kgs.~Lyngby, Denmark.}

\date{December 9, 2005}

\begin{abstract}
We show that in edge-source diffusion dynamics the integrated
concentration $N(t)$ has a universal dependence with a
characteristic time-scale $\tau=({\calA}/{\calP})^2 \pi/(4D)$,
where $D$ is the diffusion constant while $\calA$ and $\calP$ are
the cross-sectional area and perimeter of the domain,
respectively. For the short-time dynamics we find a universal
square-root asymptotic dependence $N(t)=\No \sqrt{t/\tau}$ while
in the long-time dynamics $N(t)$ saturates exponentially at $\No$.
The exponential saturation is a general feature while the
associated coefficients are weakly geometry dependent.
\end{abstract}


\maketitle


\begin{figure}[b!]
\centerline{\epsfig{file=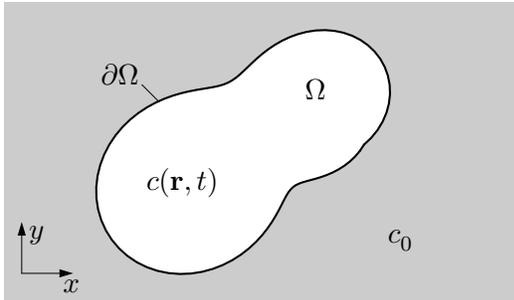,
width=0.8\columnwidth}} \caption{A sketch of the cross section
$\Omega$ (white) with boundary $\partial\Omega$ (black curve) in
the $xy$-pane of a rod with translational invariance in the $z$
direction. At time $t=0$, where the concentration
$c(\textbf{r},t)$ inside $\Omega$ is zero, diffusion is suddenly
turned on from the outside, where the concentration is kept at the
constant value $\co$ (gray). \label{fig:geometry} }
\end{figure}

Concepts like diffusion and Brownian motion are central in a wide
range of complex dynamical phenomena~\cite{Hanggi:05a,Hanggi:05b}
including diffusion of ions through biological membranes, neutron
diffusion in nuclear reactors, charge-carrier diffusion in
semiconductors, diffusion of heat in any substance, diffusion of
momentum in fluids, and diffusion of photons in the interior of
the Sun~\cite{Landau:87a,Smith:89a,Cussler:1997a,Bird:02a}. In its
simplest form of a scalar quantity $c$ the diffusion dynamics is
governed by the linear partial differential equation
 \begin{equation}\label{eq:diffusion}
 D\nabla^2c(\rrr,t) = \partial_t c(\rrr,t),
 \end{equation}
where $D$ is the diffusion constant. With the above notation we
have emphasized diffusion of matter (with concentration $c$), but
we note that the same equation also governs diffusion of energy
such as in thermal problems where Eq.~(\ref{eq:diffusion}) is
often referred to as the heat equation. Despite its apparent
simplicity the link between space and time variables $\rrr$ and
$t$ typically makes diffusion dynamics strongly dependent on the
geometry and initial conditions. In this paper we find an
exception to this and report a universality in edge-source
diffusion dynamics. Assuming perfect translation invariance along
the $z$ direction, we consider a cross section $\Omega$ in the
$xy$ plane, see Fig.~\ref{fig:geometry}, where chemical species or
heat is supplied at the boundary $\partial\Omega$. For simplicity
we imagine a situation with a constant concentration $\co$ outside
$\Omega$ while the domain itself is empty, $c(\rrr,t<0)=0$ for
$\rrr\in\Omega$, before the onset of diffusion at $t=0$.

Recently, pressure-driven flow in steady-state was analyzed in the
framework of simple geometrical measures \cite{mortensen:05b} such
as the cross-sectional area $\calA=\int_\Omega d\rrr$, the
perimeter of the boundary $\calP =\int_{\partial\Omega}d\ell$, and
the compactness $\calC=\calP^2/\calA$. In the following we study
the importance of these parameters for diffusion dynamics and in
particular for the integrated concentration $N(t)$ given by
 \begin{equation}\label{eq:N}
 N(t) = \int_\Omega d\rrr\, c(\rrr,t),
 \end{equation}
with the limits $N(t\rightarrow0)=0$ and
$N(t\rightarrow\infty)=\co \calA$ implied by the boundary and
initial conditions for $c$.

{\it Dimensional analysis}. As a first step in solving the
dynamics of the integrated concentration $N(t)$ one might estimate
the time-scale $\tau$ for filling up the domain. Obviously,
increasing the area $\calA$ results in an increasing filling time,
while increasing the perimeter $\calP$ or the diffusion constant
$D$ results in a decreasing filling time. By dimensional analysis
we thus arrive at $\tau\sim ({\calA}/{\calP})^2/D$ which, as we
shall see, is indeed a good estimate since detailed analysis
yields
 \begin{equation}\label{eq:tau1}
 \tau= \left(\frac{\calA}{\calP}\right)^2 \:\frac{\pi}{4D}.
 \end{equation}

\emph{Short-time diffusion dynamics}. On a short time-scale the
diffusion is perpendicular to the boundary and thus the problem is
quasi-one dimensional. By a short time-scale we here mean $t\ll
L_{\partial\Omega}^2/D$ where $L_{\partial\Omega}^{{}}$ is a
characteristic length scale (such as the local radius of
curvature) for shape variations along the boundary
$\partial\Omega$. We base our analysis on the well-known method of
combination of variables originally introduced by
Boltzmann~\cite{Cussler:1997a,Boltzmann:1894a}: normalization of
length scales by $\sqrt{4Dt}$ reduces Eq.~(\ref{eq:diffusion}) to
a 1D ordinary differential equation with the solution
 \begin{equation}
 \frac{c(\rrr,t)}{\co}=
 \textrm{erfc}\left(\frac{d_\perp(\rrr)}{\sqrt{4Dt}}\right),
 \end{equation}
where $d_\perp(\rrr)$ is the normal-distance to the boundary and
$\textrm{erfc}(x)$ is the complementary error function. For $t \ll
\tau$ the integrated concentration $N(t)$ becomes
 \begin{equation}\label{eq:N_shorttime}
 N(t)\approx \calP   \int_0^\infty ds\,\co
 \textrm{erfc}\Big(\frac{s}{\sqrt{4Dt}}\Big)=\No
 \sqrt{\frac{t}{\tau}},
 \end{equation}
where we have introduced the characteristic time scale defined in
Eq.~(\ref{eq:tau1}) and $N_0=c_0\calA$. The square-root dependence
is a universal property for geometries with sufficiently smooth
boundaries and dynamics deviating from this dependence is referred
to as anomalous. Previous work on the heat content in the crushed
ice model has reached results equivalent to
Eq.~(\ref{eq:N_shorttime}) for the heat
contents~\cite{vandenberg:1994a}.

\begin{table}[t!]
\caption{The parameters $\alpha^{{}}_1$ Eq.~(\ref{eq:alphaDef})
and $\beta^{{}}_1$ Eq.~(\ref{eq:betaDef}) for the lowest
eigenfunction for differently shaped cross sections.}
\label{tab:parameters}
\begin{tabular}{lccc}\hline\hline
Cross section & $\alpha^{{}}_1$& $\quad$ & $\beta^{{}}_1$\\\hline
Circle & $\frac{\pi\gamma_{0,1}^{2_{}}}{16}\simeq 1.14$ && $4/\gamma_{0,1}^{2_{}}\simeq 0.69$\\[1.5mm]
Half-circle & $\frac{\pi\gamma^{2_{}}_{1,1}}{(4+8/\pi)^2} \simeq
1.08$ && $0.64$\footnotemark[1]\footnotetext{\footnotemark[1]Data
  obtained by finite-element simulations \cite{Comsol:b}.}\\[1.5mm]
Quarter-circle & $\frac{\pi\gamma^{2_{}}_{2,1}}{(4+16/\pi)^2}\simeq
1.00$ && $0.65$\footnotemark[1]\\[1.5mm]
 Ellipse(1:2) & $1.18$\footnotemark[1]&&$0.67$\footnotemark[1]\\
 Ellipse(1:3) & $1.21$\footnotemark[1]&&$0.62$\footnotemark[1]\\
 Ellipse(1:4) & $1.23$\footnotemark[1]&&$0.58$\footnotemark[1]\\
 \hline
 Triangle(1:1:1)\footnotemark[2]\footnotetext{\footnotemark[2]See
 e.g.\ Ref.~\onlinecite{Brack:1997} for the eigenfunctions and eigenspectrum.} &
 $ \pi^3/36\simeq 0.86$&&$6/\pi^2\simeq 0.61$\\
 Triangle(1:1:$\sqrt{2}$)\footnotemark[3]\footnotetext{\footnotemark[3]See
 e.g.\ Ref.~\onlinecite{Morse:1953} for the eigenfunctions and
 eigenspectrum.} &$ \frac{5\pi^3} {16( 2 + \sqrt{2})^2}\simeq 0.83$&&$512/9\pi^4\simeq 0.58$\\
   \hline
 Square(1:1) & $ \pi^3/32\simeq 0.97$&&$64/\pi^4\simeq 0.66$\\
 Rectangle(1:2) & $ 5\pi^3/144\simeq 1.08$&&$64/\pi^4\simeq 0.66$\\
 Rectangle(1:3) & $ 5\pi^3/128\simeq 1.21$&&$64/\pi^4\simeq 0.66$\\
 Rectangle(1:4) & $ 17\pi^3/400\simeq 1.32$&&$64/\pi^4\simeq 0.66$\\
 Rectangle(1:$\infty$) & $ \sim\pi^3/16\simeq 1.94$&&$64/\pi^4\simeq 0.66$\\
 Rectangle(w:h) & $ \frac{\pi^3}{16}\:\frac{\calC -8}{\calC},
 \calC = \frac{4(h+w)^2}{hw}$&&$64/\pi^4\simeq 0.66$\\
 \hline
 Pentagon &  $1.02$\footnotemark[1]&&$0.67$\footnotemark[1]\\
 Hexagon &  $1.05$\footnotemark[1]&&$0.68$\footnotemark[1]\\
 \hline\hline
\end{tabular}
\end{table}

\emph{Long-time diffusion dynamics}. Since $N(t)\leq \No$ the
result in Eq.~(\ref{eq:N_shorttime}) is of course only meaningful
for $t\ll\tau$. When time becomes comparable to $\tau$ a
saturation will occur due to decreasing gradients in density. For
structures without high symmetries the saturation will be
accompanied by an onset of diffusion parallel to the boundary
$\partial\Omega$, and in this limit the dynamics will be slow
compared to the initial behavior, Eq.~(\ref{eq:N_shorttime}). To
study this we first derive a continuity equation by applying
Green's theorem~to~Eq.~(\ref{eq:diffusion}),
\begin{equation}\label{eq:continuity}
D\int_{\partial\Omega} d\ell\, \nnn\cdot \nablabf c(\rrr,t) =
\partial_t N(t).
\end{equation}
Here, $\nnn$ is a normal vector to $\partial\Omega$, the integral
is a line integral along $\partial\Omega$, and $-D\nablabf c$ is
naturally interpreted as a current density. Next, we note that for
long time-scales we have to a good approximation that $\nnn\cdot
\nablabf c(\rrr,t)\propto \co-N(t)/\calA$ is constant along the
boundary so that
 \begin{equation}
 \No-N(t) \propto
 \partial_t N(t),
 \end{equation}
resulting in an exponentially decaying difference. This may also
be derived from an eigenfunction expansion,
 \begin{equation}
 \label{eq:cexpand}
 \frac{c(\rrr,t)}{\co} =
 1-\sum_n f^{{}}_n\phi^{{}}_n(\rrr)e^{-\alpha^{{}}_n t/\tau},
 \end{equation}
which upon substitution into the diffusion equation yields a
Helmholz eigenvalue problem for $\phi^{{}}_n$ and $\alpha^{{}}_n$,
 \begin{equation}
 \label{eq:alphaDef}
 -\nabla^2 \phi^{{}}_n(\rrr) =
 \frac{4}{\pi}\left(\frac{\calP}{\calA}\right)^2\alpha^{{}}_n
 \phi^{{}}_n(\rrr)
 \end{equation}
with $\phi^{{}}_n(\rrr)=0$ for $\rrr \in\partial\Omega$.
Eq.~(\ref{eq:cexpand}) and the initial condition $c(\rrr,t=0)=0$
imply that $\sum_n f^{{}}_n\phi^{{}}_n(\rrr)=1$ and thus $f^{{}}_n
= \int_\Omega d\rrr\, \phi^{{}}_n(\rrr)/\int_\Omega d\rrr\,
\big|\phi^{{}}_n(\rrr)\big|^2$. The long-time dynamics is governed
by the lowest eigenvalue $\alpha^{{}}_1$ yielding
 \begin{align}
 \frac{N(t)}{\No } &=1- \sum_n\beta^{{}}_n\exp\Big(-\alpha^{{}}_n
 \frac{t}{\tau}\Big)\nonumber\\
 &\approx  1- \beta^{{}}_1\exp\Big(-\alpha^{{}}_1 \frac{t}{\tau}\Big),
 \quad t\gg\tau,
 \label{eq:Nlongtime1}
 \end{align}
where
 \begin{equation}
 \label{eq:betaDef}
 \beta^{{}}_n= \frac{\calA_n^{\rm eff}}{\calA},\quad
 \calA_n^{\rm eff} = \frac{\big|\int_\Omega d\rrr\,
 \phi^{{}}_n(\rrr)\big|^2}{\int_\Omega d\rrr\,
 \big|\phi^{{}}_n(\rrr)\big|^2}.
 \end{equation}
As often done in optics~\cite{Mortensen:02a}, $\calA_n^{\rm eff}$
can be interpreted as the effective area covered by the $n$th
eigenfunction. Values for a selection of geometries are tabulated
in Table 1. The circle is the most compact shape and consequently
it has the largest value for $\beta^{{}}_1$, or put differently
the mode has the relatively largest spatial occupation of the
total area. The normalized eigenvalue $\alpha^{{}}_1$ is of the
order unity for compact shapes and in general it tends to increase
slightly with increasing surface to area ratio $\calP/\calA$. The
modest variation in both $\alpha^{{}}_1$ and $\beta^{{}}_1$ among
the various geometries suggests that the overall dynamics of
$N(t)$ will appear almost universal and that, e.g.,
Eq.~(\ref{eq:Nlongtime1}) for the circle (mathematical details
follow below),
 \begin{equation}\label{eq:Nlongtime}
 \frac{N(t)}{\No }\approx  1- \frac{4}{\gamma^{2_{}}_{0,1}}
 \exp\Big(-\gamma^{2_{}}_{0,1}
 \frac{\pi}{16}\:\frac{t}{\tau}\Big),
 \quad \gamma^{2_{}}_{0,1}\simeq 5.783,
 \end{equation}
will account quantitatively well even for highly non-circular
cross sections.

\emph{Analytical and numerical examples}. In the following we
consider a number of geometries and compare the above asymptotic
expressions to analytical and numerical results. For the numerics
we employ time-dependent finite-element
simulations~\cite{Comsol:b} and solve Eq.~(\ref{eq:diffusion})
with a subsequent numerical evaluation of the integrated
concentration, Eq.~(\ref{eq:N}).

The circular cross section serves as a reference and an
illustrative example where we directly can compare the above
limits to analytical and numerical results. Applying the method of
separation of variables yields \cite{Cussler:1997a}
\begin{figure}[t!]
\centerline{\epsfig{file=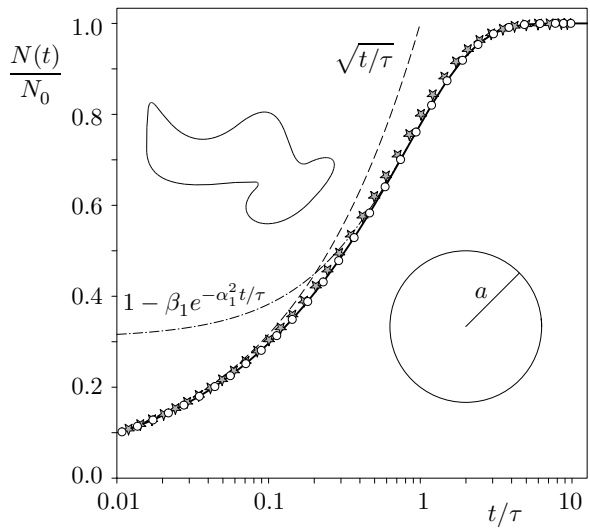,
width=0.9\columnwidth}}
\caption{A lin-log plot of $N(t)/\No$ as a
function of $t/\tau$ for a circular cross section (white circles)
and for an arbitrarily shaped cross section (gray stars); both
shapes are shown in the insets. The solid line shows the exact
result for the circle, i.e., the first 1000 terms in the infinite
series Eq.~(\ref{eq:Ncircle_exact}), the dashed line shows the
short-time asymptotic expression Eq.~(\ref{eq:N_shorttime}), the
dot-dashed line shows the long-time asymptotic expression
Eq.~(\ref{eq:Nlongtime}) for the circle with $\alpha^{{}}_1 =
\gamma^{2_{}}_{0,1}\pi/16$ and $\beta^{{}}_1 =
4/\gamma^{2_{}}_{0,1}$, while the data-points are the results of
time-dependent finite-element simulations. \label{fig:circle} }
\end{figure}
 \begin{equation}
 \frac{c_{\rm circ}(\rrr,t)}{\co}=1-2\sum_{n=1}^\infty
 \frac{J^{{}}_0(\gamma^{{}}_{0,n} r/a)}{\gamma^{{}}_{0,n}
 J^{{}}_1(\gamma^{{}}_{0,n})}\:e^{-\gamma^{2_{}}_{0,n} Dt/a^2 },
 \label{eq:cExpandCirc}
 \end{equation}
where $a$ is the radius and $\gamma^{{}}_{m,n}$ is the $n$th zero
of the $m$th Bessel function of the first kind,
$J^{{}}_m(\gamma^{{}}_{m,n})=0$. By a straightforward integration
over the cross section we get
 \begin{equation}\label{eq:Ncircle_exact}
 \frac{N_{\rm circ}(t)}{ \No}=1- \sum_{n=1}^\infty
 \frac{4}{\gamma^{2_{}}_{0,n}}\:
 \exp\bigg(-\gamma^{2_{}}_{0,n}
 \frac{\pi}{16}\:\frac{t}{\tau}\bigg),
 \end{equation}
where we have made the time-scale $\tau$ explicit, see
Eq.~(\ref{eq:tau1}). This result can also be derived using the
continuity equation, Eq.~(\ref{eq:continuity}). For $t> \tau$ the
series converges rapidly, and keeping only the first term we
arrive at Eq.~(\ref{eq:Nlongtime}).

In Fig.~\ref{fig:circle} we compare the asymptotic results,
Eqs.~(\ref{eq:N_shorttime}) and (\ref{eq:Nlongtime}) with the
exact result, Eq.~(\ref{eq:Ncircle_exact}), as well as with
time-dependent finite-element simulations. As seen both the
short-time square-root and long-time exponential dependencies are
in good agreement with the exact results as well as with the
simulations. Fig.~\ref{fig:circle} also includes numerical results
for an arbitrarily shaped cross section and, as suggested above,
we see that Eqs.~(\ref{eq:N_shorttime}) and~(\ref{eq:Nlongtime})
account remarkably well even for this highly non-circular shape.

\begin{figure}[t!]
\centerline{\epsfig{file=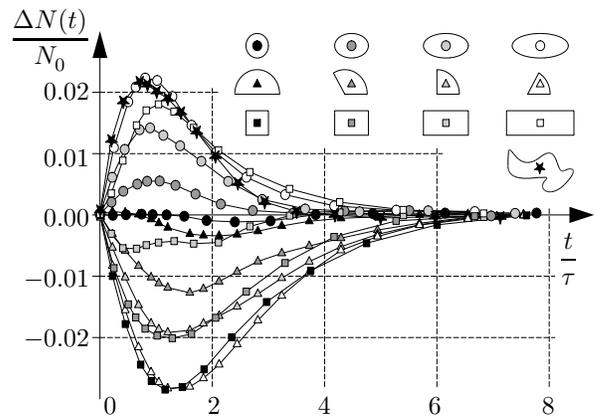,
width=0.9\columnwidth}} \caption{Plot of the deviation $\Delta
N(t)/\No$ from the circular case, Eq.~(\ref{eq:DeltaN}), as a
function of $t/\tau$ for different cross sections (see insets).
The data-points are results of finite-element simulations. The
ellipses have eccentricities $1, 1.5, 2$ and 2.5, the circle
sectors have angles $\pi, 2\pi/3, \pi/2$ and $\pi/3$, and the
rectangles have aspect ratios $1, 1.5, 2$ and $3$. Note how the
maximal deviation is less than $\pm0.03$ or $\pm 3$\%.
\label{fig:differentshapes} }
\end{figure}

In order to see how well Eq.~(\ref{eq:Ncircle_exact}) accounts for
other non-circular geometries we have employed time-dependent
finite-element simulations to numerically study the relative
deviations $\Delta N/\No$ from it,
\begin{equation}\label{eq:DeltaN}
\frac{\Delta N(t)}{\No} \equiv \frac{N(t)-N^{{}}_{\rm
circ}(t)}{\No}.
\end{equation}
Figure~\ref{fig:differentshapes} summarizes results for a number
of geometries. In all cases the dynamics at small times $t \ll
\tau$ is in full accordance with the predicted square-root
dependence, Eq.~(\ref{eq:N_shorttime}), and for long times the
predicted exponential dependence, Eq.~(\ref{eq:Nlongtime}), fits
the dynamics excellently. For the dynamics around $t\sim \tau$
deviations from the circular result are well within $\pm 3\%$ for
the considered highly non-circular geometries.

\emph{Discussion and conclusion}. We have shown that edge-source
diffusion dynamics in a rod of arbitrary cross section $\Omega$
has an intrinsic time-scale $\tau=({\calA}/{\calP})^2 \pi/(4D)$,
with $D$ being the diffusion constant while $\calA$ and $\calP$
are the cross-sectional area and perimeter of $\Omega$,
respectively. Initially, the filling $N(t)$ follows a universal
square-root dependence $N(t)=\No \sqrt{t/\tau}$, irrespectively of
the shape of the domain $\Omega$. For longer times $N(t)$
saturates exponentially at $\No$. The saturation is governed by
the lowest dimensionless eigenvalue $\alpha^{{}}_1$ of the
Helmholz equation rather than the full spectrum. Since
$\alpha^{{}}_1$ depends only weakly on the geometry the dynamics
becomes almost universal. Numerically, we have observed that the
deviation from strict universality is typically less than a few
percent.

\begin{figure}[t!]
\centerline{\epsfig{file=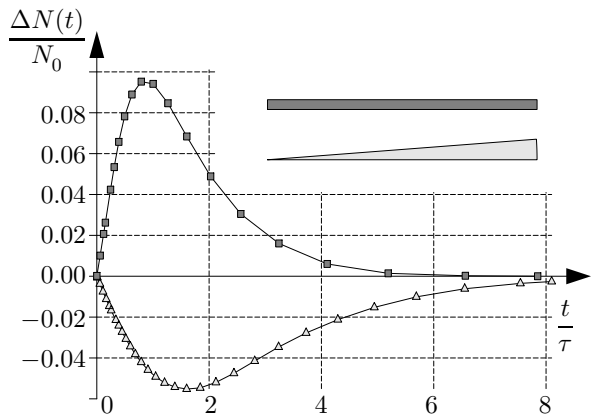,
width=0.9\columnwidth,clip}} \caption{Plots as in
Fig.~\ref{fig:differentshapes} but here for two extreme
geometries: a sector of a circular disc with angle
$2\arctan(1/30)$ and a rectangle with aspect ratio $1/30$, see
insets. While these two shapes have nearly the same area and
perimeter, the former fills slower than the circular cross section
and the latter faster. The maximal deviations are $-6$\% and
$+10\%$, respectively. \label{fig:ExtremeShapes} }
\end{figure}

The diffusion problem presented here relates to the question posed
by Mark Kac~\cite{Kac:1966}: "\emph{Can one hear the shape of a
drum}?". In the present diffusion problem knowledge about the
short-time dynamics allows one to extract the area to perimeter
ratio $\calA/\calP$ while the shape itself cannot be inferred. For
the long-time diffusion dynamics strict universality would require
that different shapes have Helmholz eigenfunctions with the same
set of eigenvalues $\big\{\alpha^{{}}_n\big\}$ (isospectrality)
and effective areas $\big\{\beta^{{}}_n\big\}$. However, since the
answer to the question of Kac in most cases is positive, see
however Refs.~\cite{Gordon:92,Sridhar:94}, the eigenfunction
properties $\big\{\alpha^{{}}_n\big\}$ and
$\big\{\beta^{{}}_n\big\}$ of different geometries do differ. It
is thus only the short-time dynamics which is strictly universal
while, as mentioned above, the long-time dynamics depends weakly
on shape through the first dimensionless eigenvalue
$\alpha^{{}}_1$ and the corresponding dimensionless effective area
$\beta^{{}}_1$.

Our simulations support these conclusions, see
Fig.~\ref{fig:differentshapes}, and even for extreme shapes such
as the narrow disc sector with angle $2\arctan(1/30)$ and the flat
rectangle with aspect ratio $1/30$ the deviations from
Eq.~(\ref{eq:Ncircle_exact}) are less than 10\% around
$t\sim\tau$, see Fig.~\ref{fig:ExtremeShapes}. These extreme
shapes have almost the same area and perimeter while the
eigenfunction properties $\big\{\alpha^{{}}_n\big\}$ and
$\big\{\beta^{{}}_n\big\}$ are very different, e.g.,
$\beta^{{}}_1$ is constant $64/\pi^4$ for all rectangles, see
Table~\ref{tab:parameters}, while it scales as $m^{-2/3}$ for the
disc sector with angle $\pi/m$. For any aspect ratio $h/w$ the
lowest eigenfunction $\phi^{{}}_1 = \sin(\pi x/w)\sin(\pi y/h)$ of
the rectangle is nearly uniformly distributed in $\Omega$, and the
shape favors rapid perpendicular diffusion resulting in a filling
slightly faster than for the circular shape. For the disc sector
with angle $\pi/m$ the lowest eigenfunction $\phi^{{}}_1 =
J^{{}}_m(\gamma^{{}}_{m,1}r/a) \sin(m\theta)$ is confined to a
region of width $a/m^{1/3}$ near the circular edge. This shortens
the effective perimeter resulting in a filling time longer than
for the circular shape. It is thus possible to find extreme shapes
where $\calA/\calP$ is no longer the characteristic length-scale
for the long-time diffusion dynamics. The same applies for a cross
section with the shape of a tear drop \cite{Weisstein:04c}, and
for more exotic geometries, such as cross sections with a fractal
polygonal boundary~\cite{Vandenberg:1999}, we expect more severe
deviations from the dynamics reported here. However, the
deviations from strict universality obtained by extending the
short-time scale $\tau$ to the long-time regime are remarkably
small.

Apart from the fascinating and intriguing physics involved we
believe our results are important to a number of practical
problems including mass diffusion in microfluidic channels and
heat diffusion in arbitrarily shaped rods.

\emph{Acknowledgements}. We thank Steen Markvorsen and Ole Hansen
for stimulating discussions and Michiel van den Berg for directing
our attention to previous work. This work is supported by the
Danish Technical Research Council (Grant
Nos.~26-03-0073~and~26-03-0037).


\end{document}